\documentclass[12pt,titlepage]{article}
\usepackage{amsmath}
\usepackage{amssymb}
\usepackage{graphicx}
\usepackage{caption2}
\usepackage{amsfonts}
\usepackage{cite}
\usepackage{lineno}

\newcommand{\be}{\begin{equation}}
\newcommand{\ee}{\end{equation}}
\newcommand{\bea}{\begin{eqnarray}}
\newcommand{\eea}{\end{eqnarray}}
\newcommand{\bef}{\begin{figure}}
\newcommand{\ef}{\end{figure}}
\newcommand{\bt}{\begin{tabular}}
\newcommand{\et}{\end{tabular}}
\newcommand{\bno}{\begin{enumerate}}
\newcommand{\eno}{\end{enumerate}}

\def\3{\ss}

\catcode`\"=\active
\def"{\accent'177}

\begin{document}

\linenumbers

\begin{center}

{\Large\bf  Two new standing  solitary  waves  in  shallow  water}\\  \vspace{0.5cm}

Shijun Liao\footnote{To whom correspondence should be addressed. sjliao@sjtu.edu.cn} \\  \vspace{0.5cm}

State Key Laboratory  of  Ocean Engineering \&   Dept. of Mathematics\\
School of Naval Architecture, Ocean and Civil Engineering\\
Shanghai Jiao Tong University,  Shanghai 200240, China

\end{center}

\hspace{-0.75cm}{\bf Abstract} {\em  In this paper,  the closed-form analytic solutions of two new Faraday's  standing  solitary  waves  due to the parametric resonance of liquid in a vessel vibrating vertically with a constant frequency are given for the first time.    Using  a model  based on  the  symmetry of wave elevation  and  the  linearized Boussinesq equation,   we  gain the  closed-form wave  elevations  of  the  two  kinds  of non-monotonically  decaying   standing  solitary  waves  with  smooth crest and the even or odd symmetry.     All of them have never been reported,  to the best of our knowledge.   Besides,  they  can  well  explain some  experimental  phenomena.     All of these are helpful to deepen and enrich our understandings about standing solitary waves and Faraday's wave.   
}

\hspace{-0.75cm} {\bf PACS Number}: 47.35.Bb

\hspace{-0.75cm} {\bf Key words}  Standing  wave,  solitary,  Faraday's wave,  parametric resonance   

\hspace{-0.75cm} \hrulefill

\section{Introduction}

As pointed out by Faraday \cite{Faraday1831} and Benjamin \& Ursell \cite{Benjamin1954},  when a vessel containing liquid  vibrates vertically with a constant driving frequency $\Omega$,  the so-called parametric resonance occurs so that  standing surface waves are observed,  in case that the liquid oscillates with a constant frequency $\omega$ that is half of the driving frequency  $\Omega$, say, $\omega=\Omega/2$.   In fact, Faraday waves have been observed in many fields of science.  For example,  
the experimental observation of Faraday waves in a Bose-Einstein condensate was reported by  Engels et al  \cite{Engels2007}, and the Faraday instability on a free surface of superfluid $^4$He was investigated by  Abe et al  \cite{Abe2007} and Ueda et al \cite{Ueda2007}.  
  In 2011, using  a  vertically  vibrating  Hele-Shaw   cell    (i.e. nearly two dimensional)  partly filled with water,   Rajchenbach, Leroux and Clamond \cite{PRL2011} did an excellent experiment and observed  two new standing solitary  surface  waves with the odd or even symmetry.   These new standing waves have an unusual characteristic:   their elevations   {\em non-monotonically}   decay  to  zero  in the horizontal direction, while    vibrating periodically in the vertical direction.  Especially,  they  pointed  out  that  ``the existence of an oscillion of odd parity had never been reported in any media up to now''.  To the best of our knowledge,    theoretical solutions have never  been found  for these new standing solitary waves.

 In this paper, the  closed-form analytic solutions of  two new Faraday's solitary  waves\footnote{Here, the solitary wave means the localized wave.}   due to the parametric resonance of liquid in a vessel vibrating vertically with a constant frequency are reported.   Using a model based on the symmetry of wave elevation and the  linearized Boussinesq equation \cite{Boussinesq1871},   we gain the  closed-form  solution  of    two  kinds  of  {\em non-monotonically}  decaying   standing  solitary  waves  with  the even or odd symmetry.    Both of them   have never been reported, to the best of our knowledge.    Especially,  they  can  well  explain  some  experimental phenomena  currently reported  by  Rajchenbach, Leroux and Clamond \cite{PRL2011}.     All of these are helpful to deepen and enrich our understandings about standing solitary waves and Faraday's wave.      

\section{Closed-form solutions of the new standing waves}

Consider a two-dimensional (2D) Faraday's  waves  in the water depth $h$,   excited  by a vertically vibrating horizontal bottom being  purely sinusoidal with a single driving frequency $\Omega$.     Let $\omega$ denote  the frequency of the excited  standing wave,   respectively.   In theory,   it is well-known  that the parametric resonance occurs when $\omega = \Omega/2$,  say,  the driving frequency $\Omega$ of the bottom is twice the frequency $\omega$ of the liquid vibration \cite{Faraday1831, Benjamin1954},   which was   currently  confirmed  once again by  the  excellent  experiment  of  Rajchenbach, Leroux and Clamond  \cite{PRL2011}.    Thus, in this paper,  we focus on the case $\omega =\Omega/2$ for the parametric resonance.    

We use here a model based on the symmetry of wave elevation and the linearized  Boussinesq  equation \cite{Boussinesq1871}.   Let $\eta(x,\tau)$ denote the dimensionless wave elevation, where  $\tau = \omega\; t = \Omega \; t/2$  denote the dimensionless time and   $x$ is the dimensionless horizontal coordinate with $x=0$ corresponding to the wave crest.   
According to the excellent experiment of Rajchenbach, Leroux and Clamond \cite{PRL2011},  we  have many reasons  to  assume that  
the wave elevation has either the even symmetry  about the wave crest $x=0$,  i.e. 
\begin{equation}
\eta(x,\tau) = \eta(-x,\tau) ,   \;\;   -\infty < x < +\infty,   \label{symmetry:even}
\end{equation}
or  the  odd symmetry 
\begin{equation}
\eta(x,\tau) = -\eta(-x,\tau),    \;\;   -\infty < x < +\infty, \label{symmetry:odd}
\end{equation}
respectively.   Assuming that the crest is smooth,  the even symmetry (\ref{symmetry:even}) gives us the boundary condition
\begin{equation}
\eta_x(0,\tau) = 0.  \label{bc:even}
\end{equation}
Besides,  the odd  symmetry (\ref{symmetry:odd})  is equivalent to the boundary condition
\begin{equation}
 \eta(0,\tau)=0. \label{bc:odd}
\end{equation}
 Using the above symmetry and the boundary condition at $x=0$,   we only need  seek  a solution $\eta(x,\tau)$ in the interval $0 <  x < +\infty$.    It should be emphasized that the symmetry plays an important role in our approach, as shown below.

Assumed that the fluid is inviscid, incompressible, and the flow is irrotational in $0<x < +\infty$  (i.e. the flow is not necessarily irrotational at $x=0$ ).   Such kind of free surfaces in the interval  $0<x < +\infty$  can be  modeled  approximately  by  the famous Boussinesq  equation \cite{Boussinesq1871}, which describes many wave phenomena  in shallow water.     In physics,  the principle of relativity requires that the equations describing the laws of physics have the same form in all admissible frames of reference.   Therefore,   following Boussinesq \cite{Boussinesq1871} and using water depth $h$ as a characteristic length,  one can gain         
the dimensionless Boussinesq equation in the reference-frame fixed with the vertically vibrating bottom: 
\begin{eqnarray}
&&\eta_{\tau\tau}  - g'  \left(\eta_{xx}+\frac{1}{3} \eta_{xxxx} +3\eta \eta_{xx} + 3\eta_x \eta_x \right) = 0, \;\; 0  <  x < +\infty,  \;\;\;   \label{eq:exact}
\end{eqnarray}
subject to the bounded  condition
\begin{equation}
|\eta(x,\tau)| <  S, \hspace{1.0cm}  0  \leq   x <+\infty,  \label{bc:infinity}
\end{equation}
where  $g'$ is  the so-called  dimensionless  ``apparent gravity acceleration'' and the $S$ is a large enough positive constant,  respectively.     Note that the above equation has exactly the same form as the traditional Boussinesq equation \cite{Boussinesq1871}, except that the ``gravity acceleration'' term $g'$ has a different meaning.   Obviously, according to  Einstein's  theory of general relativity,  for an observer moving with the vertically vibrating horizontal bottom that is not an inertial frame of reference,  the  so-called  apparent  gravity acceleration reads   \[  g' =  G \left( 1-F\cos 2\tau\right),\] where  $G = g/(h \omega^2)$  is the dimensionless acceleration of gravity, $g$  is the acceleration due to gravity,  $F = \Gamma/g$  denotes the dimensionless driving acceleration with  $\Gamma$  being  the amplitude  of the forcing acceleration of the bottom,  respectively.    

Assume that the wave amplitude is  so small  that all nonlinear terms of  (\ref{eq:exact}) can be neglected.   Thus,  we have the linearized  Boussinesq  equation in the non-inertial frame of reference fixed  with the  vertically vibrating  bottom:
\begin{equation}
\eta_{\tau\tau} -G \left( 1-F\cos 2\tau\right)\left(\eta_{xx}+\frac{1}{3} \eta_{xxxx} \right) = 0, \;\; 0 <  x <+\infty. 
\label{geq:Boussinesq:L}
\end{equation}
Our purpose is to find the solutions of Eq. (\ref{geq:Boussinesq:L}), subject to  the bounded  condition (\ref{bc:infinity})  and either the boundary condition (\ref{bc:even})   for the standing solitary waves with the even symmetry of elevation or (\ref{bc:odd}) with the odd symmetry,  which  oscillate  periodically in time $\tau$ with the period $T = 2\pi$.

Note that  $\tau = \omega t$  and $\omega=\Omega/2$,  where $\Omega$  is  the driving frequency of the vertically vibrating bottom.   According  to  the   excellent  experiment done by  Rajchenbach, Leroux and Clamond  \cite{PRL2011},  the  parametric resonance occurs when $\omega = \Omega/2$.  Thus,  we  express the standing wave in the form 
\begin{equation}  
\eta(x,\tau)  = f(\tau) \; e^{\lambda \; x},   \;\;\;   0  <  x < +\infty,   \label{eta:transform}
\end{equation}
where $f(\tau)$ is a periodic function with the period $T = 2\pi$, and $\lambda$ is an unknown  eigenvalue to be determined.    Note that the eigenvalue $\lambda$ can be   real or complex.   Substituting the above expression  into (\ref{geq:Boussinesq:L}), we have  a linear ordinary different equation
\begin{equation}
f''(\tau)-G \lambda^2 \left( 1+\frac{\lambda^2}{3}\right) (1-F \cos 2\tau) f(\tau) = 0.
\end{equation}
The above equation can be rewritten as the standard  Mathieu equation
\begin{equation}
f''(\tau) + \left[ a - 2 q \cos(2\tau)\right] f(\tau) = 0, \label{geq:Mathieu}
\end{equation}
where 
\begin{equation}
a = -G \lambda^2 \left( 1+\frac{\lambda^2}{3}\right), \;\; q = \left(\frac{F}{2} \right) a. \label{def:a,q}
\end{equation}
Thus, $f(\tau)$ is a periodic Mathieu function  with  the  characteristic value $a$ and  the  parameter $q$,  denoted by $f(\tau) = M_c(\tau; a,q)$.    Note that  similar Mathieu-type analyses  have been carried out for Faraday waves in one  and two-component  Bose-Einstein condensates \cite{Nicolin2011, Nicolin2012, Balaz2012}

It is well-known that, for a given non-zero  parameter  $q$,  the corresponding Mathieu functions $f(\tau)$ of Eq. (\ref{geq:Mathieu}) are  periodic in $\tau$ only for certain  values of $a$,  called  Mathieu characteristic values.    According to Floquet's Theorem, any Mathieu function $f(\tau)$ can be written in the form $e^{i r \tau} f^*(\tau)$, where $f^*(\tau)$ has period $2\pi$ and $r$ is the Mathieu characteristic exponent.  The Mathieu function $f(\tau)$ is periodic only when the characteristic exponent $r$ is an integer or rational number.  

For given characteristic value $a$ and parameter $q = (a F) /2$,  let us consider the even Mathieu function $f(\tau)$  of  (\ref{geq:Mathieu}) with the characteristic exponent $r=1$ so that $f(\tau)$ has the period $T=2\pi$.   As mentioned above, the characteristic value of the even Mathieu function $f(\tau)$  with characteristic exponent $r=1$    given by the parameter $q$  must be equal to the characteristic value  $a$ itself.     This gives,  by means of the computer algebra system Mathematica,   the  following nonlinear  algebraic  equation
\begin{equation}
\mbox{MathieuCharacteristicA[1, $a F/2$]} = a,  \label{geq:a}
\end{equation}
where the Mathematica command {MathieuCharacteristicA}[$r$, $q$] is used  to gain the characteristic value $a$ for even Mathieu functions with characteristic exponent $r$ and the given parameter  $q$.   Given the dimensionless driving acceleration $F$,  the above nonlinear algebraic equation contains only the unknown characteristic value  $a$, denoted by $a^*$.      The corresponding solution \[ f(\tau) = M_c(\tau; a^*, F a^*/2)\]  is an even Mathieu function with the period $2\pi$.    It should be emphasized that the  characteristic  value  $a^*$ depends only on the dimensionless driving acceleration $F$.  

\begin{figure}[thbp]
\centering
\includegraphics[scale=0.45]{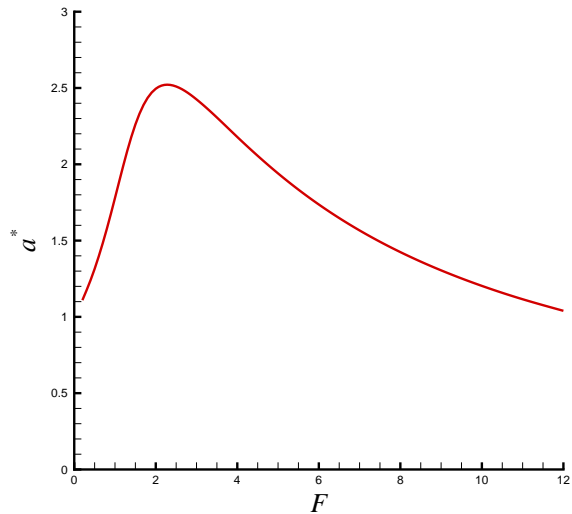}
\caption{ The positive characteristic  values of $a^*$ versus the dimensionless driving acceleration $F$ given by the linearized Boussinesq equation (\ref{geq:Boussinesq:L}). }
\label{figure:a-positive}
\end{figure}

\begin{figure}[thbp]
\centering
\includegraphics[scale=0.45]{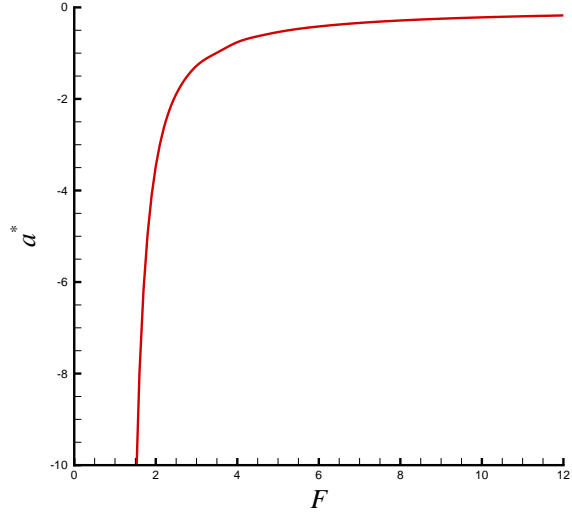}
\caption{ The negative characteristic values of $a^*$ versus the dimensionless driving acceleration $F$ given by the linearized Boussinesq equation (\ref{geq:Boussinesq:L}). }
\label{figure:a-negative}
\end{figure}

It is found that the nonlinear algebraic equation (\ref{geq:a}) has two solutions in general.  For example,  when $F=2$, we have a positive characteristic value $a^*= 2.49527$ and a negative characteristic value $a^*= -3.47044$, respectively.   For different values of the dimensionless driving acceleration $F$,  we have different characteristic value $a^*$.   The two curves of the characteristic value $a^*$  versus  the dimensionless driving acceleration  $F$ are as shown in Figs. \ref{figure:a-positive} and \ref{figure:a-negative}.    Note that,  the maximum positive characteristic  value $a^*_{max}$ is $2.52168$,  corresponding to the dimensionless driving acceleration  $F = 2.28$.    Note that, according to the excellent experiment of Rajchenbach, Leroux and Clamond \cite{PRL2011},   the  parametric resonance was found when $F = 2.0$.   So,  the above theoretical result can partly explain why  Rajchenbach, Leroux and Clamond  \cite{PRL2011}  observed the parametric resonance in case of $F=2.0$,  since the corresponding  characteristic  value $a^*=2.49527$  is  rather close to $a^*_{max}= 2.52168$ so that the parametric resonance is easier to be created and observed.  We will discuss this later in details.

According to (\ref{def:a,q}),  as long as  the characteristic value $a^*$ is known, it is easy to  gain the unknown eigenvalue $\lambda$ by solving the nonlinear algebraic equation
\begin{equation}
\lambda^2 \left( 1+\frac{\lambda^2}{3}\right)+\mu = 0,
\end{equation}  
where
\[  \mu=\frac{a^*}{G}. \]
Thus,  $\lambda$ is dependent upon the dimensionless gravity acceleration $G$ and the dimensionless driving acceleration $F$, since $a^*$ is determined by $F$ only.  When $\mu > 3/4$,  the above nonlinear algebraic equation have four complex roots
\begin{equation}
\lambda_{1,2,3,4} = \pm \sqrt{-\frac{3}{2} \pm i  \sqrt{3\left(\mu -\frac{3}{4}\right)}},
\end{equation} 
where $i=\sqrt{-1}$ denotes the imaginary  unit.  When $0<\mu < 3/4$, there exist four pure  imaginary  roots
\begin{equation}
\lambda_{1,2}=\pm  i \sqrt{\frac{3}{2} \pm \sqrt{3\left(\frac{3}{4}-\mu \right)}}.
\end{equation}
When $\mu<0$, there are  two pure imaginary  roots 
\begin{equation}
\lambda_{3,4} = \pm i \sqrt{\frac{3}{2}\left(\sqrt{1-\frac{4}{3}\mu } +1\right)}
\end{equation}
and  two real roots
\begin{equation}
\lambda_{3,4} = \pm \sqrt{\frac{3}{2}\left(\sqrt{1-\frac{4}{3} \mu }-1 \right)}.
\end{equation}

Thus,  when $\mu>3/4$,  we have  four complex roots  $\lambda=\pm \alpha \pm \beta i$ with $\alpha>0$ and $\beta >0$, 
corresponding to the wave elevation in a general form 
\begin{eqnarray}
\eta(x,\tau) &=&  A_1 \; Mc(\tau; a^*,q^*)   e^{- \alpha x} \left( \cos\beta x  + i \sin\beta x \right)\nonumber\\
&+& A_2 \; Mc(\tau; a^*,q^*)   e^{- \alpha x} \left( \cos\beta x  - i \sin\beta x \right)\nonumber\\
&+& A_3 \; Mc(\tau; a^*,q^*)   e^{+ \alpha x} \left( \cos\beta x  + i \sin\beta x \right)\nonumber\\
&+& A_4 \; Mc(\tau; a^*,q^*)   e^{+ \alpha x} \left( \cos\beta x  - i \sin\beta x \right),
\end{eqnarray}
where $A_1, A_2, A_3$ and $A_4$ are constants.  
However,  restricted by  the bounded condition (\ref{bc:infinity}),  only the solution in the form 
\begin{eqnarray}
\eta(x,\tau) &=& A_1 \; Mc(\tau; a^*,q^*)   e^{- \alpha x} \left( \cos\beta x  + i \sin\beta x \right)\nonumber\\
&&  + A_2 \; Mc(\tau; a^*,q^*)   e^{- \alpha x} \left( \cos\beta x  - i \sin\beta x \right) \nonumber\\
&=&  Mc(\tau; a^*,q^*)   e^{- \alpha x} \left( A_0 \cos\beta x  + B_0 \sin\beta x \right), \;\; 0  <  x < +\infty, 
\end{eqnarray}
has physical meanings,  where $A_0  =  A_1  +  A_2$   and  $B_0  =  (A_1-A_2)  i$  are  real  constants.  

Thus, using the boundary condition (\ref{bc:odd}) for the odd symmetry and enforcing  $A_0=0$,   we have  the  wave elevation
\begin{eqnarray}
\eta(x,\tau) =B_0 \; Mc(\tau; a^*,q^*) \sin(\beta x)  e^{ - \alpha  x  } , \hspace{1.0cm} 0  <  x< +\infty.  \label{eta:exp:sin:x>0}
\end{eqnarray}  
Then, due to the odd symmetry (\ref{symmetry:odd}),  we  have  the   odd-pattern  elevation  
\begin{eqnarray}
\eta(x,\tau) =B_0 \; Mc(\tau; a^*,q^*) \sin(\beta x)  e^{ - \alpha  |x| } , \hspace{1.0cm} -\infty <  x< +\infty.  \label{eta:exp:sin}
\end{eqnarray} 
Similarly,  using the boundary condition (\ref{bc:even}) for the even symmetry (\ref{symmetry:even}),  we have  the even-pattern wave elevation
\begin{eqnarray}
\eta(x,\tau) &=&A_0 \; Mc(\tau; a^*,q^*)   e^{-|\alpha x|}   \left[ \cos(\beta x)  + \left( \frac{\alpha}{\beta} \right)\; \sin(\beta |x|) \right],    \label{eta:exp:cos:sin}
\end{eqnarray}
which has a smooth crest and is valid in the whole domain $ -\infty<x<+\infty$.   Note that the  standing solitary wave elevations (\ref{eta:exp:sin})  and  (\ref{eta:exp:cos:sin}) decay  non-monotonically in the $x$ direction, and have no peaked crest.

 \begin{figure}[thbp]
\centering 
\includegraphics[scale=0.35]{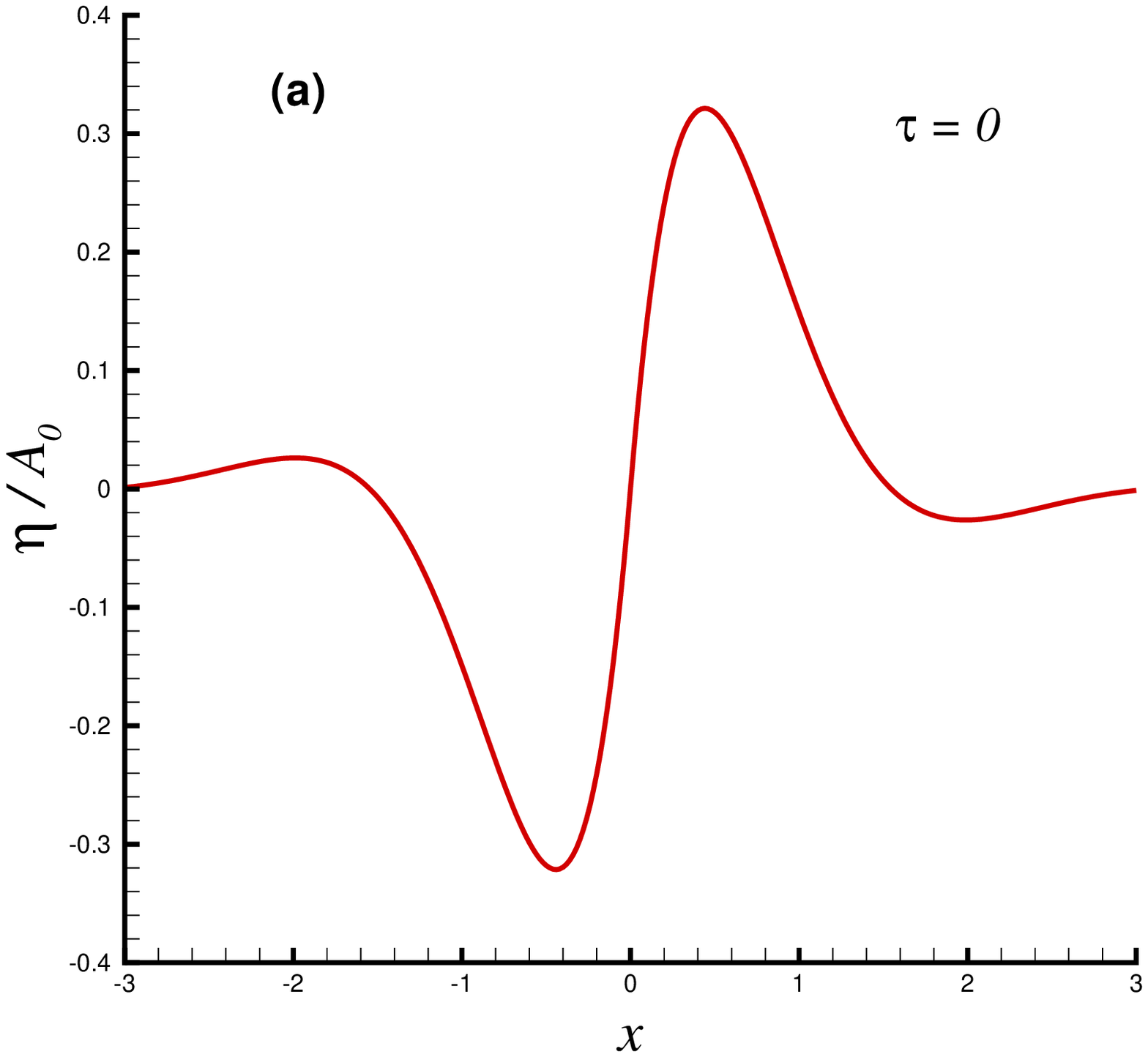}
\includegraphics[scale=0.35]{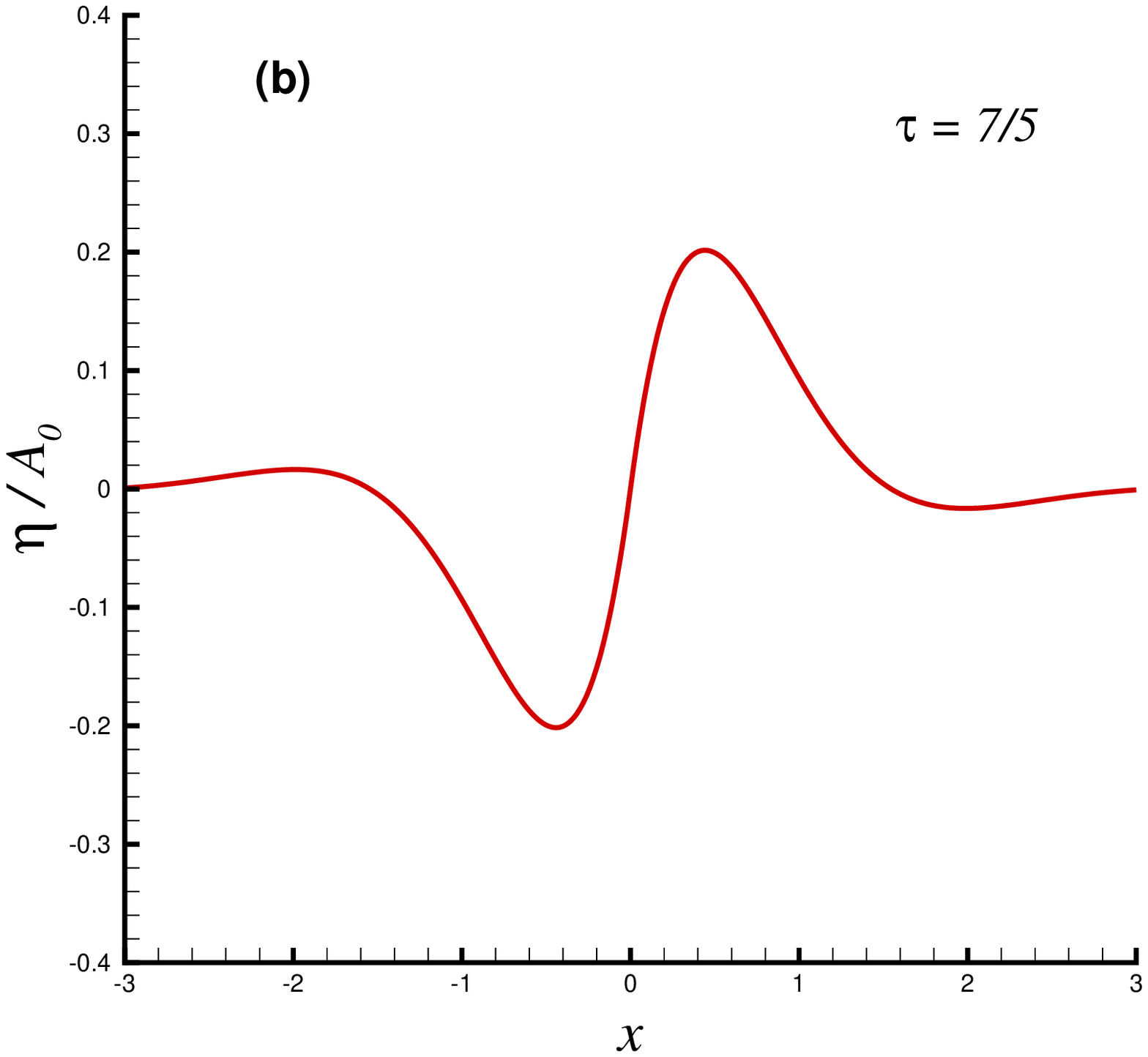}
\includegraphics[scale=0.35]{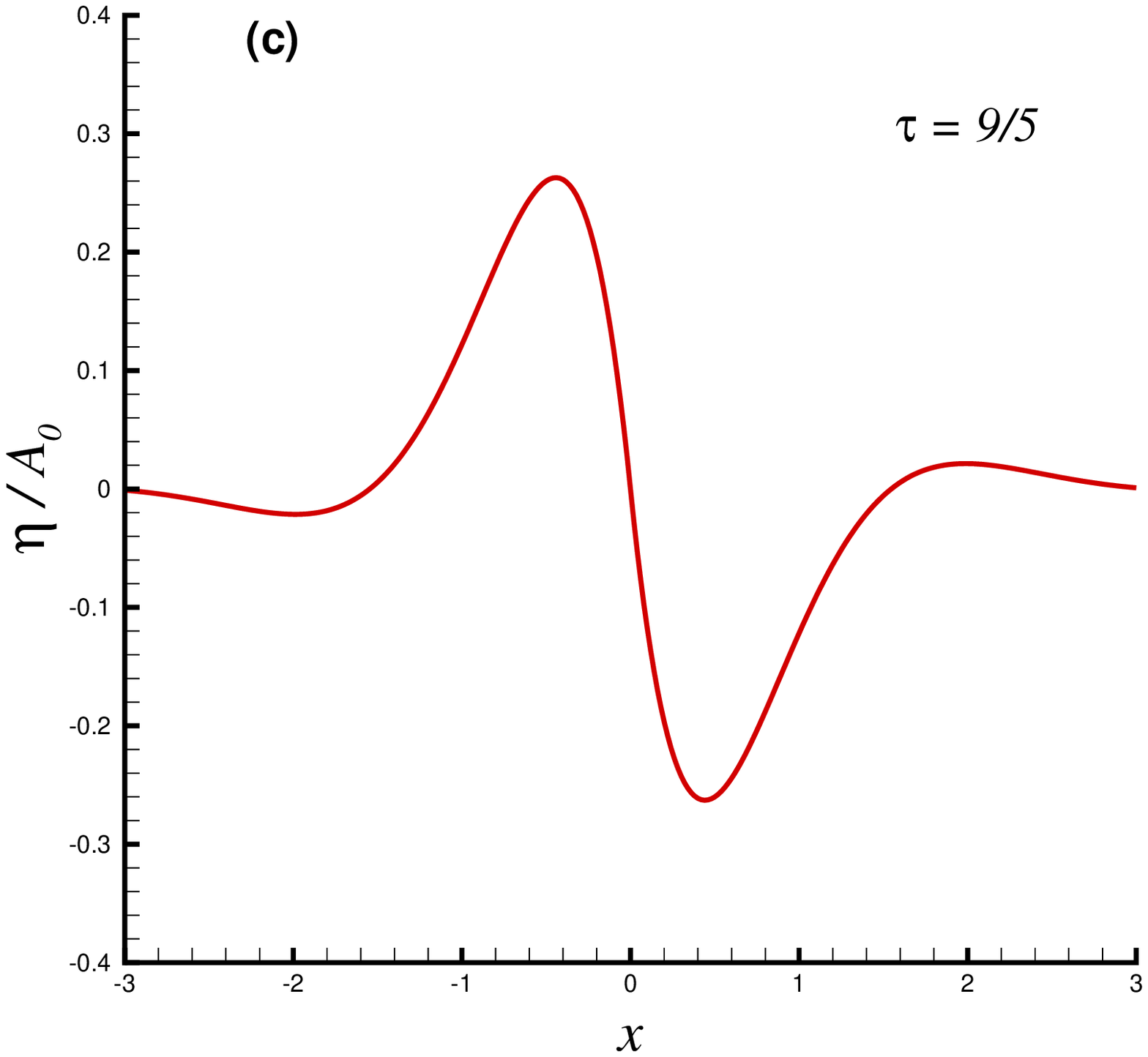}
\includegraphics[scale=0.35]{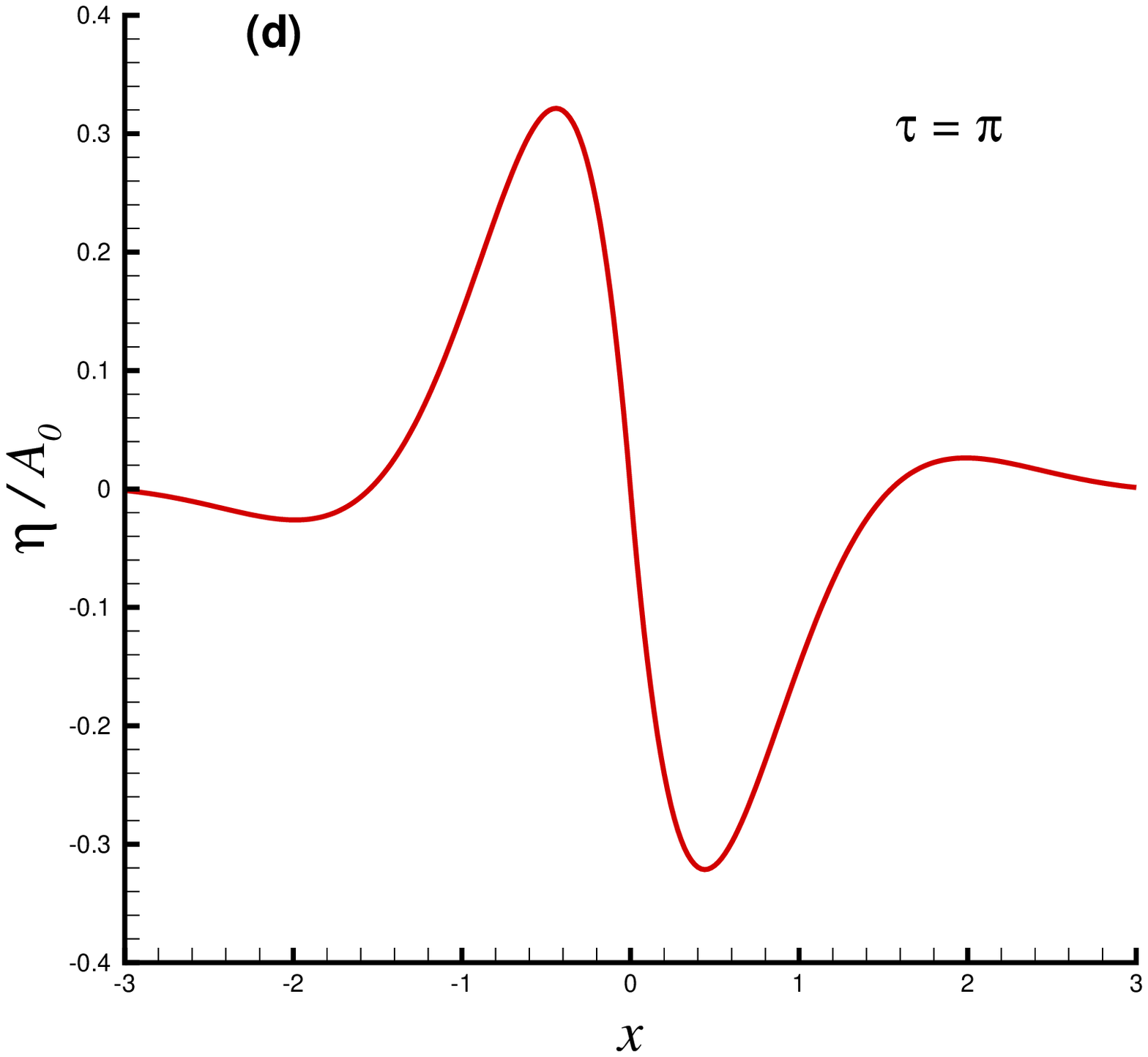}
 \caption{Non-monotonically  decaying   standing  solitary  wave (\ref{eta:exp:sin:example}) with the odd symmetry and a smooth crest.  (a): $\tau = 0$;  (b): $\tau=7/5$;  (c): $\tau =9/5$; (d):  $\tau = \pi$. }
\label{figure:eta:exp:sin}
\end{figure}

 \begin{figure}[thbp]
\centering
\includegraphics[scale=0.45]{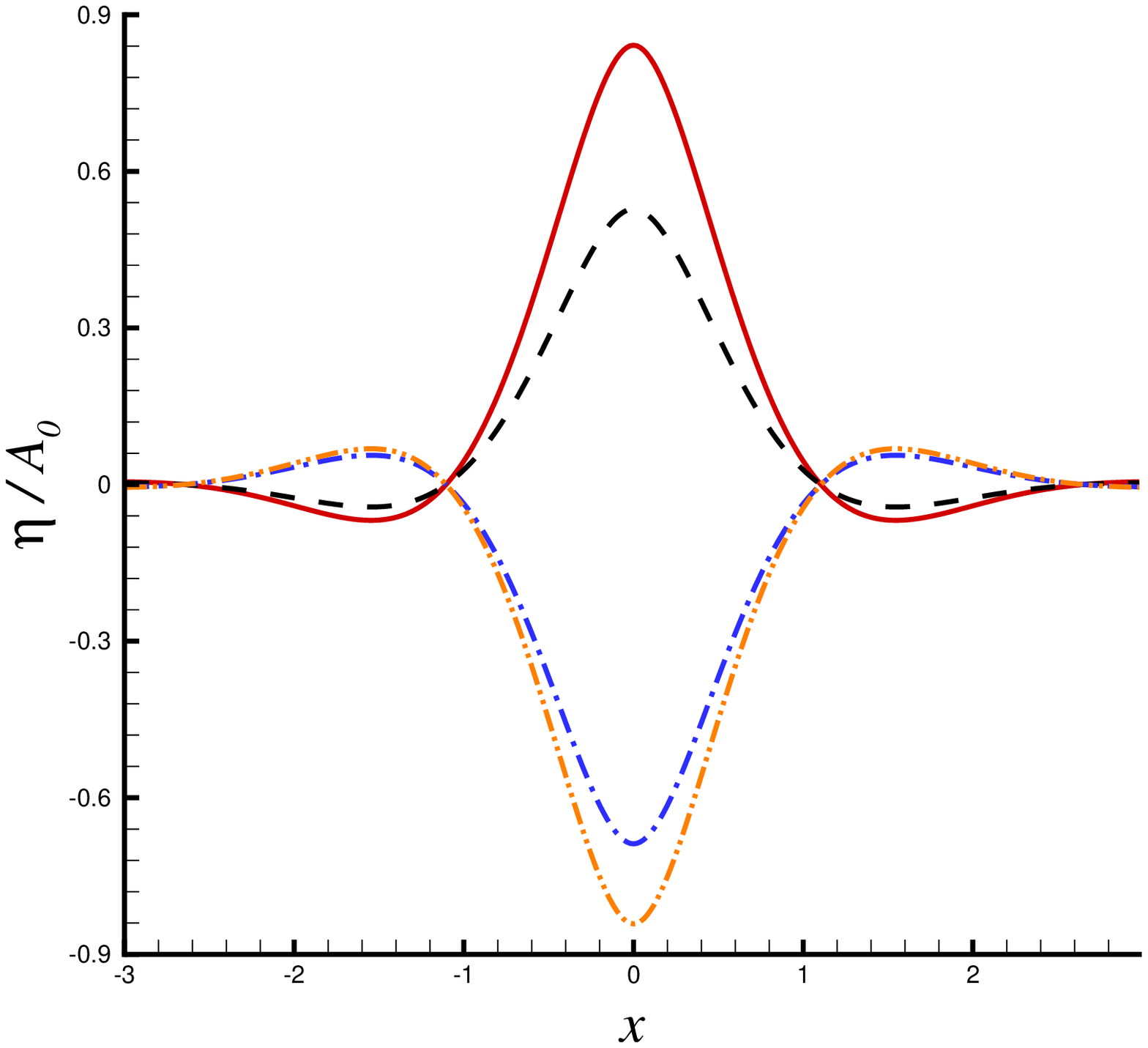}
\caption{Non-monotonically  decaying   standing  solitary  wave (\ref{eta:exp:cos:sin:example}) with the even symmetry and a smooth crest.  Solid line: $\tau = 0$;  Dashed line: $\tau=7/5$; Dash-dotted line: $\tau =3\pi/4$; Dash-dot-dotted line:  $\tau = \pi$.  }
\label{figure:eta:exp:cos:sin}
\end{figure}

For example, in case  of  the driving frequency $\Omega=11$ Hz with the vibration amplitude 4.1 mm and water depth 5 cm, which were used by  Rajchenbach, Leroux and Clamond \cite{PRL2011} in their excellent experiment,  we have the dimensionless driving acceleration 
$F \approx  2$ and the dimensionless gravity acceleration  $G \approx 0.164$.   When $F =2$, there exist  one positive characteristic  $a^*= 2.49527$ and one negative characteristic  $a^*= -3.47044$, corresponding to $\mu = a^*/G = 15.2151$ and $\mu=-21.1612$, respectively.    Especially,  when $a^* = 2.49527$,  i.e.  $\mu= 15.2151$,   we have four complex eigenvalues 
\[  \lambda = \pm  1.62113 \pm 2.03126 i, \]
corresponding to  a non-monotonically decaying standing solitary wave with  the odd symmetry  and  the smooth crest 
\begin{eqnarray}
\eta(x,\tau) &=&  A_0 M_c(\tau;2.49527,2.49527)  \sin(2.03126 x) \;  e^{-1.62113 |x|}, \label{eta:exp:sin:example}
\end{eqnarray} 
as shown in Fig.~\ref{figure:eta:exp:sin},  and the wave elevation with even symmetry and smooth crest 
 \begin{eqnarray}
\eta(x,\tau) &=&  A_0 M_c(\tau;2.49527,2.49527)  \;  e^{-1.62113 |x|} \nonumber\\
&& \times \left[   \cos\left(2.03126 x\right) +0.798091 \sin\left(2.03126 |x|\right)\right], \label{eta:exp:cos:sin:example}
\end{eqnarray} 
as shown in Fig.~\ref{figure:eta:exp:cos:sin}, respectively.  

It should be emphasized that the standing solitary wave (\ref{eta:exp:sin:example})  has the odd parity about $x=0$.    Note that  Rajchenbach, Leroux and Clamond \cite{PRL2011}  found  a similar standing solitary wave with the odd parity in their excellent experiment,  and  
  pointed  out  that  ``the existence of an oscillon of odd parity had never been reported in any media up to now''.    Thus,  the closed-form solution (\ref{eta:exp:sin:example}) might provide a theoretical explanation  for this experimental phenomenon.   

Note that  the standing solitary  waves (\ref{eta:exp:sin:example}) and (\ref{eta:exp:cos:sin:example}) do {\em not} decay  monotonically,  as shown in Figs.~\ref{figure:eta:exp:sin} and \ref{figure:eta:exp:cos:sin},  which are  qualitatively  similar  to  those experimentally found by Rajchenbach, Leroux and Clamond \cite{PRL2011}.      Note that these   non-monotonically decaying  standing solitary waves  (\ref{eta:exp:sin:example}) and (\ref{eta:exp:cos:sin:example})  are not exactly the same as those found by the excellent experiment of Rajchenbach, Leroux and Clamond \cite{PRL2011}.   Such kind of difference  may  likely  be  attributed  to  the  probe  motion in their experiment \cite{PRL2011}, and also to the neglect of the nonlinearity of the Boussinesq equation that is valid for  fairly  long  waves  with  small-amplitude in shallow water.   The nonlinearity of the Boussinesq equation might affect the eigenvalue $\lambda$,    which determines the decay-rate of wave elevation.  However,  the  nonlinear  terms  should not  qualltatively  influence the shape of wave elevation.   Since we mainly focus on the shape of wave elevation in this article,  the neglect of the nonlinear terms is acceptable.       Obviously,  better analytic approximations of the two new standing solitary waves should be gained, if the exact Boussinesq equation (\ref{eq:exact}) or the fully nonlinear wave equation is solved.   

\section{Discussions and concluding remarks}

 In this paper,  some new Faraday's waves  due to the parametric resonance of liquid in a vessel vibrating vertically with a constant frequency are reported.   Using a model based on the symmetry of wave elevation and the linearized  Boussinesq equation,   we gain the  closed-form  solutions  of   two  kinds  of  {\em non-monotonically}  decaying   standing  solitary  waves  with  the  even  or  odd  symmetry.    They can  well explain, although partly, some  experimental phenomena  currently reported by  Rajchenbach, Leroux and Clamond \cite{PRL2011}. 
 
 First,  our  closed-form solution (\ref{eta:exp:sin})  has the odd parity about $x=0$, as shown in Fig.~\ref{figure:eta:exp:sin}.    Note that  Rajchenbach, Leroux and Clamond \cite{PRL2011}  found  a similar standing solitary wave with the odd parity in their excellent experiment,  and  
 pointed  out  that  ``the existence of an oscillon of odd parity had never been reported in any media up to now''.  So,  our  closed-form solution (\ref{eta:exp:sin})  supports   this experimental phenomenon.         

Secondly, based on the linearized  Boussinesq  equation, the characteristic value $a^*$ is dependent upon the dimensionless driving acceleration $F$ only.  Thus,  for given dimensionless gravity acceleration $G=g/(h \omega^2)=4g/(h \Omega^2)$,  the occurrence of the non-monotonically decaying  standing solitary waves  mainly depends on the dimensionless driving acceleration $F = \Gamma/g$ of the vertically vibrating bottom:  the larger $a^*$, the larger possibility of  the occurrence of the non-monotonically  decaying  standing  solitary  waves,  since $\mu = a^*/G > 3/4$ is the criterion for the linearized Boussinesq equation.    According to Fig.~\ref{figure:a-positive},   the  non-monotonically decaying  standing  solitary  waves  occur   with  the   maximum  possibility  at   $F \approx  2.28$,  corresponding to the maximum characteristic value $a^*_{max} = 2.52168$.   Note that   Rajchenbach, Leroux and Clamond \cite{PRL2011}  observed  the two non-monotonically  decaying  standing solitary waves at $F = 2$, corresponding to the characteristic value $a^*=2.49527$ that is rather close to $a^*_{max} = 2.52168$.   So, our theoretical result can explain this  experimental phenomena quite well.  

Thirdly,  Rajchenbach, Leroux and Clamond  \cite{PRL2011} found experimentally that  the two non-monotonically decaying  standing solitary  waves  occur in an interval  $F _L < F < F_R$.  They gave it a theoretical explanation using Meron's  stability theory \cite{Meron1987}.   Based  on  the  linearized  Boussinesq  equation,  we gain the criterion of occurrence of the two non-monotonically decaying  standing solitary waves: $\mu  = a^*/G  > 3/4$.   According to Fig.~\ref{figure:a-positive}, $a^*$ has a maximum $a^*_{max} = 2.52168$  at $F = 2.28$.   So,  given a proper value of $G$,  one might find a closed interval of $F$ for the occurrence of the two  non-monotonically  decaying  standing solitary waves.   Thus, our theoretical  result  about the  criterion $\mu>3/4$  can  explain this experimental phenomenon, too.   

Seriously speaking,  profile of the  standing waves  should be dependent upon not only the dimensionless accelerations $G, F$
 but also  the wave height.   However,  based on the linearized Boussinesq equation,  the profile of the two standing waves is dependent on   $G$ and $F$ only.   This is  similar to the periodic travelling waves, whose wave profile is  sinusoidal  and  independent of wave height.   Besides, the detailed evolution of the standing waves should be also influenced by wave height.    So, if the influence of wave height is considered,   the nonlinear Boussinesq equation or even the fully nonlinear water wave equations should be used for  more accurate wave profile  and better understandings about such kind of standing waves.   
 
Note that,using the even or odd symmetry, we have  either the boundary condition (\ref{bc:even}) or (\ref{bc:odd})  at $x=0$,  so that  it is {\em enough} for the governing equation to be  satisfied in the interval $0 <  x  < +\infty$ {\em except} $x=0$.    This  is  well-known and  widely  used  in the theory of differential equations.       

 Traditionally, one need give a global expression of a solution in the whole domain.  However, this is difficult in many cases.   Fortunately,  this traditional idea is out of date.  In modern mathematics, we often express a smooth function by a lots of  local simple functions in a finite number of sub-domains:  this idea is widely used in Finite Element Method (FEM).   Although there exists singularity at each boundary of the sub-domain where the solution is not smooth,  such kind of solutions are widely accepted and used.   In this paper,  we use different base functions to express the solution of a new type of standing waves found by experiment in two sub-domains only,  i.e. $(-\infty, 0)$  and $(0, +\infty)$, and connect them by the symmetry and the smoothness condition at  $x = 0$.    Comparing to the FEM,  the mathematical approach used in this article is more traditional: we use the symmetry to divide the whole domain into only two sub-domains, and besides the solutions are smooth at  $x = 0$.  
 
Finally,  it should be emphasized that the closed-form wave elevations of   the  two  non-monotonically decaying  standing solitary waves (\ref{eta:exp:sin}) and (\ref{eta:exp:cos:sin})  are obtained under the assumption of the even or odd symmetry of wave elevation  by means of  the linearized  Boussinesq  equation with the neglect of  viscocity of fluid  in the interval $0<x<+\infty$.    The  symmetry  has  an  important  role  in  our  approach.  The fact that the two closed-form solutions  well  explain some phenomenon of the excellent experiment of Rajchenbach et al  \cite{PRL2011} indicates the validity of this model.    In addition,  since  the linearized  Boussinesq  equation is only a simplified model for shallow water waves,   all  conclusions and theoretical predictions  reported in  this  article  should  be further  checked  and verified  by  fine  numerical simulations  and  physical  experiments in future, even though our closed-form solutions well explain some experimental phenomena of Rajchenbach et al  \cite{PRL2011}.   All of these are helpful to deepen and enrich our understandings about standing solitary waves and Faraday's wave.       

\vspace{0.75cm}
 
\hspace{-0.75cm}{\bf Acknowledgement}  Thanks to Prof. C.C. Mei (MIT, USA) for the discussions and to the reviewers for their valuable comments.    This work is partly supported by State Key Lab of Ocean Engineering (Approval No. GKZD010056-6) and National Natural Science Foundation of China (Approval No. 11272209).   

\bibliographystyle{unsrt}

\end{document}